\documentclass[aps,prd,twocolumn,superscriptaddress,nofootinbib,10pt]{revtex4-1}

\usepackage{amsmath,amssymb,mathrsfs,graphicx,bm}
\usepackage{xcolor}
\usepackage[hidelinks]{hyperref}
\usepackage{appendix}
\usepackage{orcidlink}

\newcommand{\dd}{\mathrm{d}}
\newcommand{\ee}{\mathrm{e}}
\newcommand{\MS}{Misner--Sharp }
\newcommand{\varrhoeff}{\rho}
\newcommand{\Lin}{L_{\rm in}}
\newcommand{\Lout}{L_{\rm out}}
\newcommand{\din}{\delta_{\rm in}}
\newcommand{\dout}{\delta_{\rm out}}

\begin{document}

\title{Integrable singularities obstruct Cauchy horizons and mass inflation}

\author{Francisco Tello-Ortiz \orcidlink{0000-0002-7104-5746}}
\email{francisco.tello@ufrontera.cl}
\affiliation{Departamento de Ciencias Físicas, Universidad de La Frontera, Casilla 54-D, 4811186 Temuco, Chile.}


\begin{abstract}
Black-hole interiors admitting an inner (Cauchy) horizon are dynamically unstable through the Poisson--Israel--Ori mass-inflation mechanism, an instability governed by the Dray--'t Hooft--Redmount relation and widely regarded as a dynamical realization of strong cosmic censorship. We show that a specific class of anisotropic interiors -- static, spherically symmetric sources supported by a marginally divergent but integrable energy density $\rho\sim A/r^2$, with $A>1/(8\pi)$ and mild screening at finite radius -- is kinematically incompatible with this mechanism. For such sources the compactness $q(r)=m(r)/r$ decreases monotonically from a finite, super-critical value at the center to $1/2$ at the (unique) event horizon, producing a strictly positive \emph{compactness gap} throughout the trapped region and excluding any second zero of the lapse. We then show that this static gap has dynamical consequences in which it enters quantitatively. (i) Under the null energy condition and the dynamical form of the screening condition, $2m/r$ is non-decreasing along both future null directions of a trapped double-null domain, so the gap propagates under nonlinear spherically symmetric evolution and the lapse cannot approach zero there. (ii) For ingoing radiation, described exactly by a generalized Vaidya geometry, the gap bounds the blueshift: the advanced time spent in the gapped region is {finite}, whereas it diverges at a Cauchy horizon. (iii) For counter-streaming null dust on the fluid background, the radiative part $\mu$ of the \MS mass obeys exactly $\partial_u\partial_v\mu=2\,\partial_u\mu\,\partial_v\mu/(r|f|)$, of which the Dray--'t Hooft--Redmount relation is the thin-shell form; the coupling between ingoing and outgoing fluxes that drives mass inflation is therefore $2/(r|f|)$, and on the interior annulus it is uniformly bounded precisely when the gap is positive. This yields an explicit a priori bound on the mass for boundary data below a threshold set by the gap and, in the Ori thin-shell model, an e-folding rate bounded by the influx luminosity instead of tending to the inner-horizon surface gravity $\kappa_-$. Marginally screened integrable-singularity interiors thus realize a class of sources for which the Poisson--Israel--Ori amplification channel is structurally inoperative in the null-dust and thin-shell models in which mass inflation was originally established.
\end{abstract}

\maketitle

\section{Introduction}

Black-hole interiors in general relativity are expected to be dynamically unstable due to the phenomenon of mass inflation \cite{PoissonIsrael1989,PoissonIsrael1990,Ori1991}. In spacetimes admitting a Cauchy horizon, ingoing perturbations undergo an unbounded blueshift governed by the Dray--'t Hooft--Redmount (DTR) relation for colliding null shells \cite{DrayTHooft1985,Redmount1985}, leading to exponential growth of the \MS mass and to the breakdown of the classical interior geometry. This mechanism is widely regarded as a dynamical realization of strong cosmic censorship \cite{Dafermos2003,LukOh2019,DafermosLuk2025,DiasReallSantos2018}.

Most analyses of this instability assume the existence of an inner horizon and study its nonlinear fate under perturbations. A complementary, less explored question is whether the formation of such a horizon can be obstructed at a structural level by the matter content itself. Several works have instead modified the black-hole interior through effective or semiclassical considerations, including regular black-hole models that can retain mass inflation even in the absence of a Cauchy horizon \cite{CarballoRubio2022,CarballoRubio2024PRL}.

Here we follow a different, more modest route: rather than demanding full regularity, we consider interiors that admit an \emph{integrable singularity} at the center -- curvature invariants diverge, yet suitable volume-weighted integrals remain finite, in the sense of \cite{LukashStrokov2013} and as realized in explicit static models by \cite{Ovalle2023,Casadio2022IJMPD,CasadioGiustiOvalle2022,Casadio2023PLB,Gallerani2025} {and in the stationary case by \cite{CasadioGiustiOvalle2023}}. Whether this integrability property itself survives under generic, non-spherical perturbations is a separate and still-debated question, recently challenged on the grounds of test-field backreaction \cite{ArrecheaLiberatiNeshatVellucci2025}; this is distinct from the obstruction to mass inflation studied below, which depends only on the existence of a compactness gap in the trapped region away from the center, not on the exact preservation of integrability at $r=0$ under generic perturbations. We do not construct these interiors as candidate descriptions of realistic collapse; instead, we ask a narrower structural question: once such a matter content is assumed, under what conditions is the standard Poisson--Israel-Ori channel for mass inflation kinematically inoperative? Density profiles of marginal form $\rho\sim r^{-2}$ and the associated linear behavior of the \MS mass have been discussed before in a different context, in connection with anomalies of quasilocal mass constructions \cite{Faraoni2021,Hayward1996}. Our contribution is not the existence of such profiles, but the identification of the single quantity through which they act on the mass-inflation mechanism. The lapse $|f|=2m/r-1$ is simultaneously (a) the blueshift factor relating the advanced time of the exterior to affine time along outgoing rays, $\dd v/\dd r=2/f$, and (b) the denominator of the nonlinear coupling between ingoing and outgoing fluxes, of which the DTR relation is the thin-shell form. A uniform lower bound on $|f|$ therefore controls both ingredients of the Poisson--Israel--Ori mechanism, and we show that such a bound is preserved by the dynamics (Theorems~1 and~2 below).

The logical structure of the result is the following chain, which we state here so that the role of each hypothesis is clear before the derivation:
\begin{equation}
\begin{aligned}
&\rho\le\tfrac13\bar\rho\, \quad \mbox{and}\quad \text{NEC}
\;\Longrightarrow\;
|f|\ge\Delta_\varepsilon\ \text{(propagated)}\\
&\Longrightarrow\;
\left|\frac{\dd v}{\dd r}\right|\le\frac{2}{\Delta_\varepsilon},\quad
\frac{2}{r|f|}\le\frac{2}{r_1\Delta_\varepsilon}\\
&\Longrightarrow\;
\text{no exponential inflation.}
\label{causal_chain}
\end{aligned}
\end{equation}
The first arrow is Proposition~1 (static) and Theorem~1 (dynamical). The second arrow is where the gap is actually used, and constitutes Theorem~2: the gap bounds the blueshift and the in--out coupling, and both bounds fail precisely when $\Delta_\varepsilon\to0$. The last arrow is made quantitative in the models in which mass inflation was originally established \cite{PoissonIsrael1990,Ori1991}: ingoing radiation (generalized Vaidya), continuous counter-streaming null dust, and an outgoing thin shell crossed by an influx (the Ori model). All dynamical statements are made in spherical symmetry and for matter obeying the null energy condition (NEC).

\section{Static set-up and integrable singularities}

We consider static, spherically symmetric interiors in Schwarzschild gauge,
\begin{equation}
\dd s^2 = -f(r)\,\dd t^2 + \frac{\dd r^2}{f(r)} + r^2\dd\sigma^2,
\label{metric_static}
\end{equation}
where $\dd\sigma^2=\dd\theta^2+\sin^2\theta\,\dd\varphi^2$ is the unit-sphere line element, supported by an anisotropic fluid
\begin{equation}
T^\mu{}_\nu = \mathrm{diag}(-\rho,p_r,p_t,p_t).
\label{Tmunu_static}
\end{equation}
In this gauge, Einstein equations imply
\begin{align}
8\pi\rho &= \frac{1-f-r f'}{r^2},\label{ein_rho}\\
8\pi p_r &= -\frac{1-f-r f'}{r^2},\label{ein_pr}\\
8\pi p_t &= \frac{f'}{r}+\frac{f''}{2},\label{ein_pt}
\end{align}
so that $p_r=-\rho$ identically -- a direct consequence of staticity plus spherical symmetry in this gauge, not an additional assumption. Introducing the \MS mass via
\begin{equation}
f(r) = 1 - \frac{2m(r)}{r},
\label{MS_def}
\end{equation}
Eq.~\eqref{ein_rho} becomes
\begin{equation}
m'(r) = 4\pi r^2 \rho(r),\qquad
m(r) = 4\pi \int_0^r \rho(\bar r)\,\bar r^2\,\dd\bar r.
\label{mass_eq}
\end{equation}
The quantity controlling the interior causal structure is the compactness
\begin{equation}
q(r) \equiv \frac{m(r)}{r},\qquad f(r) = 1 - 2q(r),
\label{q_def}
\end{equation}
whose derivative, using \eqref{mass_eq} and the averaged density $\bar\rho(r)\equiv 3m(r)/(4\pi r^3)$, is
\begin{equation}
q'(r) = 4\pi r\left[\rho(r) - \frac{\bar\rho(r)}{3}\right].
\label{qprime2}
\end{equation}
The sign of $q'(r)$ is thus fixed by the comparison between the local and the averaged density -- the physical content of everything that follows.

We adapt the notion of an integrable singularity by requiring curvature invariants to diverge at the origin while the volume-weighted integral of the Ricci scalar stays finite,
\begin{equation}
\int_0^{{r_c}} r^2 |R(r)|\,\dd r < \infty
\label{integrable_def}
\end{equation}
for some ${r_c}>0$, i.e.\ the Ricci scalar is locally integrable with respect to the volume measure, in the spirit of the classical theory of distributions \cite{Hormander2005} and its use in the analysis of wave equations on curved backgrounds \cite{Friedlander1975}. For the stress tensor \eqref{Tmunu_static}, $R=-8\pi T$ with $T=-2\rho+2p_t$, and the conservation law $\nabla_\mu T^\mu{}_\nu=0$ gives $p_t=-\rho-r\rho'/2$, hence
\begin{equation}
R(r) = 8\pi\left[4\rho(r)+r\,\rho'(r)\right].
\label{R_rho}
\end{equation}
Consider a marginal profile
\begin{equation}
\rho(r) = \frac{A}{r^2} + o(r^{-2}),\qquad A>0,
\label{marginal_profile}
\end{equation}
where we write $r^2\rho(r)=A+\eta(r)$ with $\eta\in C^1$, $\eta(r)\to0$ and $r\eta'(r)\to0$ as $r\to0^+$, i.e.\ the correction is differentiably subleading; this is what makes the profile ``marginal'' rather than merely asymptotic. Then $r^2R(r)=8\pi[2A+2\eta(r)+r\eta'(r)]\to16\pi A$, so $R(r)\simeq16\pi A/r^2$ diverges, but $r^2 R(r)$ is bounded near the origin and
\begin{equation}
\int_0^{{r_c}} r^2 R(r)\,\dd r \sim \int_0^{{r_c}} \dd r
\label{ricci_conv_integral}
\end{equation}
converges, so the singularity is integrable in the sense of \eqref{integrable_def}; the precise statement is given in Appendix~A. At the same time Eq.~\eqref{mass_eq} gives $m(r)=4\pi A\,r+o(r)$, so the central compactness
\begin{equation}
q(0^+) = \lim_{r\to0^+} \frac{m(r)}{r} = 4\pi A
\label{q0_limit}
\end{equation}
is finite. If $A>1/(8\pi)$, the center already lies in the trapped region, $f(0^+)=1-8\pi A<0$. The exact power law $\rho=A/r^2$, for which $p_t=0$ and $f=1-8\pi A$ is constant, is Letelier's cloud of strings \cite{Letelier1979}; the interiors considered here are screened versions of it.

\section{Non-local criterion and compactness gap}

An inner (Cauchy) horizon would correspond to a radius $r_-\in(0,h)$ with $f(r_-)=0$, in addition to the event horizon at $r=h$; equivalently, a second zero of $q(r)-1/2$. We show that a broad class of marginally screened profiles forbids this. Assume:

(i) near the origin, $\rho(r)=A/r^2+o(r^{-2})$ with $A>1/(8\pi)$, so $q(0^+)=4\pi A>1/2$;

(ii) for all $r\in(0,h]$, $\rho(r)\le \bar\rho(r)/3$;

(iii) there exists $h>0$ with $q(h)=1/2$, where $h$ denotes the outermost zero of $f$ (the event horizon of the, static asymptotically flat geometry), and this horizon is non-degenerate: $\rho(h)<\bar\rho(h)/3$, i.e. its surface gravity $\kappa_+=f'(h)/2=-q'(h)$ is positive.

It is worth clarifying the status of hypothesis (ii), since unlike (i) and (iii) it is not an asymptotic or boundary statement but a condition imposed throughout the trapped region. A mere monotonic decrease of $\rho(r)$ would only guaranty $\rho(r)\le\bar\rho(r)$: when comparing $\rho(r)$ with the mean of $\rho$ over $[0,r]$ weighted by $\bar r^2\,\dd\bar r$, a decreasing integrand always satisfies this weaker bound. Hypothesis (ii) demands the sharper inequality $\rho(r)\le\bar\rho(r)/3$, i.e., that the density not only decrease but decrease \emph{fast enough} that its local value stays below one third of the enclosed average, a genuinely stronger requirement than monotonicity alone, and the one that turns $q'(r)\le0$ into a strict, self-sustaining property rather than a generic feature of centrally condensed matter. We do not derive (ii) from a standard energy condition or a known equation of state, and we do not claim it is generic; we regard it as a geometric sufficient hypothesis, to be checked directly on the density profile in each case, as we do below for two structurally different choices of $\rho(r)$. That said, it is not an arbitrary threshold: for the marginal profile \eqref{marginal_profile} itself, $\rho(r)\to A/r^2$ and $\bar\rho(r)\to 3A/r^2$ as $r\to0^+$, so $\rho(r)/\bar\rho(r)\to1/3$ exactly at the center. Hypothesis (ii) is therefore saturated asymptotically by the very profile the construction is built on, rather than imposed as an unrelated external constraint on top of it. Its dynamical meaning is made explicit by Theorem~1: (ii) is the static restriction of the condition under which the compactness $2m/r$ cannot decrease towards the future inside a trapped region.

\textbf{Proposition 1} (static compactness gap). \textit{Under (i)--(iii), $f<0$ on $(0,h)$, $r=h$ is the unique zero of $f$, and for every $\varepsilon\in(0,h)$
\begin{equation}
\Delta_\varepsilon \equiv
\inf_{0<r\le h-\varepsilon}\left[\frac{2m(r)}{r}-1\right]
=2q(h-\varepsilon)-1>0 ,
\label{gap_def}
\end{equation}
with $\Delta_\varepsilon=2\kappa_+\varepsilon+O(\varepsilon^2)$ as $\varepsilon\to0$.}

\textit{Proof.} By \eqref{qprime2} and (ii), $q'\le0$ on $(0,h]$, so $q(r)\ge q(h-\varepsilon)\ge q(h)=1/2$ for $0<r\le h-\varepsilon$, and the infimum in \eqref{gap_def} is attained at $r=h-\varepsilon$. If $q(h-\varepsilon)=1/2$, monotonicity would force $q\equiv1/2$ on $[h-\varepsilon,h]$ and hence $q'(h)=0$, contradicting (iii). Thus $q>1/2$ on $(0,h)$; since $h$ is the outermost zero of $f$, it is the only one. The expansion follows from $f(h-\varepsilon)=-f'(h)\varepsilon+O(\varepsilon^2)$. $\hfill\square$

Hence
\begin{equation}
f(r)=1-2q(r)\le-\Delta_\varepsilon<0\qquad(0<r\le h-\varepsilon),
\label{f_negative}
\end{equation}
and no Cauchy horizon is present: the future boundary of the trapped region is the spacelike locus $r=0$. Two remarks on the role of the hypotheses are in order. First, without (ii) the first zero of $f$ met from the center need not be the event horizon: for an asymptotically flat profile with $q(0^+)>1/2$ the zeros of $f$ come in odd number, and three zeros $h_1<h_2<h_3$ would make $h_2$ an inner horizon of the trapped shell $(h_2,h_3)$; (ii) is what excludes this. Second, the gap \eqref{gap_def} is defined, by construction, on the interval $(0,h-\varepsilon]$, which stays a finite distance away from the event horizon itself; the bound \eqref{f_negative} therefore never applies at $r=h$, where $f$ genuinely vanishes. Note also that $(0,h-\varepsilon]$ is bounded but not closed at the origin, so we refer to it as a \emph{bounded} interior region rather than a compact one throughout. The trapped region thus stays uniformly separated from $f=0$ away from the event horizon.

\begin{figure}[t]
\centering
\includegraphics[width=0.95\linewidth]{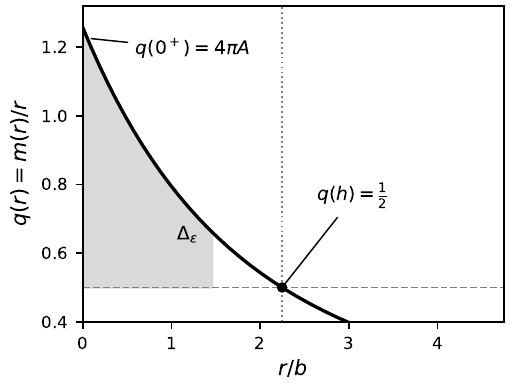}
\caption{Compactness $q(r)$ for the exponentially screened marginal profile \eqref{q_exp}, with $4\pi A>1/2$. The curve decreases monotonically from the super-critical central value $q(0^+)=4\pi A$ to $q(h)=1/2$ at the unique event horizon $r=h$. The shaded region shows the compactness gap on $(0,h-\varepsilon]$, where $q(r)-1/2\ge\Delta_\varepsilon/2$: $q(r)$ never returns to $1/2$ away from $r=h$, so no second (Cauchy) horizon forms.}
\label{fig:gap}
\end{figure}

Figure~\ref{fig:gap} illustrates this for the exponentially screened example of the next paragraph: $q(r)$ descends monotonically and uniformly stays above $1/2$ until it reaches the event horizon, never approaching it again.

\section{Example: exponentially screened marginal profile}

Consider $\rho_{\exp}(r)=(A/r^2)\ee^{-r/b}$, $A,b>0$. Then $m_{\exp}(r)=4\pi Ab(1-\ee^{-r/b})$ and, with $x\equiv r/b$,
\begin{equation}
q_{\exp}(r) = 4\pi A\,\frac{1-\ee^{-x}}{x}.
\label{q_exp}
\end{equation}
Since $\ee^{-x}(x+1)<1$ for $x>0$,
\begin{equation}
q'_{\exp}(r) = \frac{4\pi A}{b}\,\frac{\ee^{-x}(x+1)-1}{x^2}<0\qquad(r>0),
\label{qprime_exp_negative}
\end{equation}
so $q_{\exp}$ decreases strictly from $4\pi A$ at the center to zero at infinity, $q_{\exp}(h)=1/2$ defines a unique, non-degenerate event horizon, there is no inner horizon, and $\Delta^{(\exp)}_\varepsilon=2q_{\exp}(h-\varepsilon)-1>0$ for every $\varepsilon>0$. Similar conclusions hold for other screened marginal profiles (e.g. rational screeners; see Appendix~B), indicating that the compactness gap is a robust feature of this class of interiors rather than a fine-tuned property of \eqref{q_exp}.

\section{The standard mass-inflation mechanism}

The standard scenario is most transparent in double-null coordinates $(u,v)$,
\begin{equation}
\dd s^2=-\Omega^2(u,v)\,\dd u\,\dd v+r^2(u,v)\,\dd\sigma^2,
\label{doublenullmetric}
\end{equation}
with the \MS mass defined by
\begin{equation}
1-\frac{2m(u,v)}{r}= -\frac{4}{\Omega^2}\,\partial_u r\,\partial_v r.
\label{MSdouble}
\end{equation}
In Reissner--Nordstr\"om interiors, the inner horizon $r_-$ is a second zero of the lapse, $f\simeq-2\kappa_-(r-r_-)$ with $\kappa_-=|f'(r_-)|/2$ the inner-horizon surface gravity. Three facts, all traceable to this zero, make up the mechanism \cite{PoissonIsrael1990,Ori1991}; we state them quantitatively because the obstruction below acts on each of them.

\emph{(1) Infinite advanced time.} Let $v$ be the Eddington advanced time of the stationary exterior, in which the interior metric reads $\dd s^2=-f\,\dd v^2+2\,\dd v\,\dd r+r^2\dd\sigma^2$. Outgoing radial null rays obey $\dd v/\dd r=2/f$, and $r$ is an affine parameter along them. Near $r_-$ this integrates to $r-r_-\propto\ee^{-\kappa_-v}$: the whole exterior history $v\to\infty$ is received within a finite affine interval.

\emph{(2) Bounded influx, unbounded blueshift.} In this gauge the late-time influx is governed by radiative tails, $\Lin(v)\equiv4\pi r^2T_{vv}\propto v^{-p}$ \cite{Price1972,GundlachPricePullin1994}; it is bounded, and in fact decays. Boundedness of $T_{vv}$ in the exterior advanced time is therefore the standard physical input of the mass-inflation analyses, not a restriction that removes the instability. The divergence appears only in affinely normalized frames: for the tangent $k^\mu$ of an outgoing ray with $k^r=-1$, $T_{\mu\nu}k^\mu k^\nu=4T_{vv}/f^2$, which grows like $v^{-p}\ee^{2\kappa_-v}$.

\emph{(3) Nonlinear coupling to the outflux.} An influx alone produces no mass inflation; the growth of $m$ requires an outgoing component $T_{uu}$ \cite{PoissonIsrael1989,PoissonIsrael1990}. For an ingoing and an outgoing thin null shell crossing at radius $r_0$, the DTR relation \cite{DrayTHooft1985,Redmount1985,BarrabesIsrael1991} reads $f_Nf_S=f_Ef_W$ at $r_0$, where $S$ ($N$) is the sector to the past (future) of both shells and $E$ ($W$) the sector crossed by the ingoing (outgoing) shell only. In terms of the shell masses $\din\equiv m_E-m_S$ and $\dout\equiv m_W-m_S$ it is equivalent to
\begin{equation}
m_N=m_S+\din+\dout-\frac{2\,\din\dout}{r_0\,f_S(r_0)} .
\label{DTRmass}
\end{equation}
Inside the trapped region $f_S<0$ and the last term is a positive amplification. When the crossing takes place at $r_0\to r_-$, fact (1) gives $|f_S(r_0)|\propto\ee^{-\kappa_-v}$ and the amplification grows like $\ee^{\kappa_-v}$ -- the hallmark of mass inflation, $m\sim v^{-p}\ee^{\kappa_-v}$.

It is essential that this degeneracy of $f$ occurs \emph{inside} the trapped region, at a locus distinct from the event horizon, where $f$ also vanishes but for an unrelated reason (the outer boundary of trapping, not a Cauchy horizon). The compactness-gap bound \eqref{gap_def} is stated precisely on $(0,h-\varepsilon]$ for this reason: it excludes lapse degeneracy in the interior of the trapped region while saying nothing about, and not needing to say anything about, the lapse at $r=h$ itself.

Recently, \cite{CarballoRubio2022,CarballoRubio2024PRL} showed that mass inflation can persist even without a Cauchy horizon, through slowly evolving trapping horizons: the absence of an inner horizon at one instant is not by itself sufficient to prevent unbounded mass growth. Complementing this, \cite{BonannoPanassitiSaueressig2025} recently showed, for several specific regular-black-hole models (Hayward, Bardeen, Dymnikova, and an asymptotically-safe collapse model), that the growth of the \MS mass at the inner horizon is generically tamed -- power-law or even logarithmic rather than exponential -- via a dynamical basin-of-attraction analysis performed model by model; see also \cite{Visser2024} for a broader perspective on the quasi-local, rather than strictly teleological, status of the horizons involved. Our result is complementary to both: rather than solving the nonlinear dynamics for each model individually, we identify a static sufficient condition -- the compactness gap -- under which exponential growth is excluded for an entire class of interiors at once. The point raised by \cite{CarballoRubio2022,CarballoRubio2024PRL} is nevertheless decisive for what must be proven: a static gap is not enough, one must show that the lapse cannot approach zero dynamically. This is the content of Theorem~1.

\section{Dynamical obstruction to mass inflation}

We work in the double-null gauge \eqref{doublenullmetric} with $u$ and $v$ future directed, so that a sphere is trapped when $\partial_ur<0$ and $\partial_vr<0$; there $f=1-2m/r<0$ by \eqref{MSdouble}. Einstein's equations give (Appendix~C)
\begin{align}
\partial_v m &= \frac{8\pi r^2}{\Omega^2}\left(T_{uv}\,\partial_v r-T_{vv}\,\partial_u r\right),\label{massfluxeq}\\
\partial_u m &= \frac{8\pi r^2}{\Omega^2}\left(T_{uv}\,\partial_u r-T_{uu}\,\partial_v r\right).\label{massfluxeq_u}
\end{align}
In the dynamical spacetime the density is defined invariantly as $\rho\equiv2T_{uv}/\Omega^2$. For the sources of this paper, $T^a{}_b=-\rho\,\delta^a{}_b$ in the $(u,v)$ block, this is the energy density of \eqref{Tmunu_static} itself, and null dust does not contribute to it (for a general anisotropic fluid it is half the difference between the energy density and the radial pressure). We also define $\bar\rho\equiv3m/(4\pi r^3)$ as before, now with the dynamical mass. Equations \eqref{massfluxeq}--\eqref{massfluxeq_u} are equivalent to the identities
\begin{align}
\partial_v\!\left(\frac{2m}{r}\right)&=8\pi r\left(\frac{\bar\rho}{3}-\varrhoeff\right)(-\partial_v r)+\frac{16\pi r}{\Omega^2}\,T_{vv}\,(-\partial_u r),\label{mono_v}\\
\partial_u\!\left(\frac{2m}{r}\right)&=8\pi r\left(\frac{\bar\rho}{3}-\varrhoeff\right)(-\partial_u r)+\frac{16\pi r}{\Omega^2}\,T_{uu}\,(-\partial_v r).\label{mono_u}
\end{align}
On the static solutions, where $T_{uu}=T_{vv}=0$, they reduce to \eqref{qprime2}.

\textbf{Theorem 1} (propagation of the compactness gap). \textit{Let $\mathcal Q=[u_0,u_1]\times[v_0,v_1]$ be a double-null rectangle of trapped spheres in a spherically symmetric solution of Einstein's equations, and assume that on $\mathcal Q$: (a) $T_{uu}\ge0$ and $T_{vv}\ge0$ (null energy condition along the radial null directions); (b) $\rho\le\bar\rho/3$ (dynamical screening). Then $2m/r$ is non-decreasing in both $u$ and $v$ on $\mathcal Q$. In particular, if $|f|\ge\Delta$ on the past boundary $\{u=u_0\}\cup\{v=v_0\}$ of $\mathcal Q$, then $|f|\ge\Delta$ on all of $\mathcal Q$: the lapse cannot approach zero towards the future within $\mathcal Q$, and no inner horizon forms there.}

\textit{Proof.} In a trapped region $-\partial_ur>0$ and $-\partial_vr>0$, so under (a) and (b) every term on the right-hand sides of \eqref{mono_v}--\eqref{mono_u} is non-negative. Any point of $\mathcal Q$ is reached from the past boundary along a segment of constant $u$ or constant $v$, along which $|f|=2m/r-1$ does not decrease. $\hfill\square$

Theorem~1 is nonlinear in the sense that it is formulated for the full spherically symmetric Einstein equations, without linearizing the metric or the \MS mass around the static background, and uses no property of the matter beyond (a) and (b). Hypothesis (b) is the same inequality as the static hypothesis (ii), now required of the evolving density and mass, and (a) is where the outgoing component $T_{uu}$ first enters: positive outflux can only widen the gap. Reissner--Nordstr\"om illustrates the sharpness of (b): it is a source of the same type, with $\rho=Q^2/(8\pi r^4)$ and $m=M-Q^2/(2r)$, for which $\rho\le\bar\rho/3$ holds only for $r\ge Q^2/M$. The inner horizon $r_-<Q^2/M$ lies in the region where the screening condition is violated; there $2m/r$ decreases towards the future and the gap closes.

To quantify the mechanism we now specify the perturbation as in \cite{PoissonIsrael1990,Ori1991}: the static fluid of Proposition~1 plus non-interacting ingoing and outgoing null dust, $T_{vv}=\Lin(v)/(4\pi r^2)$ and $T_{uu}=\Lout(u)/(4\pi r^2)$, i.e.\ $T_{\mu\nu}=T^{(0)}_{\mu\nu}+T^{\rm in}_{\mu\nu}+T^{\rm out}_{\mu\nu}$ with each of the three components separately conserved. Equations \eqref{massfluxeq}--\eqref{mono_u} are identities valid for any spherically symmetric matter and Theorem~1 requires only its two hypotheses, whereas the equations below are specific to this model. For a fluid with $T^a{}_b=-\rho\,\delta^a{}_b$ in the $(u,v)$ block, conservation forces $\dd\rho\wedge\dd r=0$: the density remains the same function $\rho(r)$ of the areal radius in the dynamical spacetime. We may therefore split
\begin{equation}
m(u,v)=m_0\big(r(u,v)\big)+\mu(u,v),
\label{mu_def}
\end{equation}
with $m_0(r)$ the static profile \eqref{mass_eq} and $\mu$ the \emph{radiative} part of the mass; $\mu\equiv0$ is the integrable member of the family, and $\mu=\text{const}\neq0$ adds a Schwarzschild-like term to it. With ingoing flux only, the exact solution is the generalized Vaidya metric \cite{Husain1996,WangWu1999}
\begin{equation}
\dd s^2=-\Big[f_0(r)-\frac{2\mu(v)}{r}\Big]\dd v^2+2\,\dd v\,\dd r+r^2\dd\sigma^2,
\label{genVaidya}
\end{equation}
with $f_0=1-2m_0/r$, $\dot\mu=\Lin(v)$, and $v$ the advanced time of fact (1).

\textbf{Theorem 2} (the gap bounds the blueshift and the in--out coupling). \textit{Consider the static fluid of Proposition~1 perturbed by null dust obeying the NEC, $\Lin\ge0$, $\Lout\ge0$, with non-negative radiative mass on the past characteristic boundary of the trapped domain considered. Fix $0<r_1<h-\varepsilon$. Then:}

\textit{(a) The gap persists, $|f|\ge\Delta_\varepsilon+2\mu/r\ge\Delta_\varepsilon$ on $0<r\le h-\varepsilon$. For ingoing flux only [metric \eqref{genVaidya}], every future-directed causal curve in this region satisfies
\begin{equation}
\dd v\le\frac{2\,|\dd r|}{|f|}\le\frac{2\,|\dd r|}{\Delta_\varepsilon},
\label{blueshiftbound}
\end{equation}
with equality in the first relation for outgoing radial null rays. The advanced time spent in the annulus $r_1\le r\le h-\varepsilon$ is at most $2(h-\varepsilon-r_1)/\Delta_\varepsilon$, and the advanced time elapsed between $r=h-\varepsilon$ and $r=0$ is at most $2(h-\varepsilon)/\Delta_\varepsilon$. On the static background, the outgoing radial null tangent of fact (2), normalized by $k^r=-1$, obeys $T_{\mu\nu}k^\mu k^\nu=4T_{vv}/f_0^2\le4T_{vv}/\Delta_\varepsilon^2$.}

\textit{(b) Within this model, the radiative mass obeys the exact equation
\begin{equation}
\partial_u\partial_v\mu=\frac{2}{r\,|f|}\;\partial_u\mu\;\partial_v\mu ,
\label{mu_wave}
\end{equation}
with $\partial_u\mu\ge0$, $\partial_v\mu\ge0$. On every double-null rectangle $[u_0,u]\times[v_0,v]$ of trapped spheres contained in the annulus $r_1\le r\le h-\varepsilon$, the coupling coefficient satisfies $2/(r|f|)\le c_\varepsilon\equiv2/(r_1\Delta_\varepsilon)$, and
\begin{equation}
\ee^{-c_\varepsilon\mu(u,v)}\ge\ee^{-c_\varepsilon\mu(u,v_0)}+\ee^{-c_\varepsilon\mu(u_0,v)}-\ee^{-c_\varepsilon\mu(u_0,v_0)} .
\label{apriori}
\end{equation}
Denoting by $B(u,v)$ the right-hand side, the radiative mass is bounded by the data injected through the past boundary,
\begin{equation}
\mu(u,v)\le-\frac{1}{c_\varepsilon}\ln B(u,v)\qquad\text{whenever }B(u,v)>0,
\label{muboundB}
\end{equation}
which holds in particular when $\mu(u_0,v_0)=0$ and both $\mu(u,v_0)$ and $\mu(u_0,v)$ are smaller than $(\ln2)/c_\varepsilon$.}

\textit{(c) For two thin shells crossing at $r_0\in[r_1,h-\varepsilon]$, the DTR amplification in \eqref{DTRmass} is $2\din\dout/(r_0|f_S(r_0)|)\le c_\varepsilon\,\din\dout$. For a continuous influx crossing one outgoing thin shell of mass $\delta(v)$ at $r=R(v)$ in the geometry \eqref{genVaidya} (the Ori model \cite{Ori1991}), along the portion of its trajectory with $r_1\le R(v)\le h-\varepsilon$,
\begin{equation}
\frac{\dd\ln\delta}{\dd v}=\Gamma(v)\equiv\frac{2\,\Lin(v)}{R\,|f(v,R)|}\le c_\varepsilon\,\Lin(v),
\label{Gamma_def}
\end{equation}
so that, if $\Lin\le L_{\max}$,
\begin{equation}
\ln\frac{\delta(v)}{\delta(v_i)}\le c_\varepsilon\!\int_{v_i}^{v}\!\Lin\,\dd s\le\frac{4L_{\max}(h-\varepsilon)}{r_1\Delta_\varepsilon^2} .
\label{deltabound}
\end{equation}}

\textit{Proof.} (a) Since $\partial_u\mu,\partial_v\mu\ge0$ in a trapped region by \eqref{massfluxeq}--\eqref{massfluxeq_u} and the NEC, $\mu\ge0$ persists, and $|f|=|f_0(r)|+2\mu/r\ge\Delta_\varepsilon$ by \eqref{f_negative}; this is Theorem~1 in the present model, where (b) of that theorem holds because the density is still the static profile and $m\ge m_0$, so that $\rho(r)\le\bar\rho_0/3\le\bar\rho/3$ with $\bar\rho_0\equiv3m_0/(4\pi r^3)$. For \eqref{genVaidya}, a future-directed causal displacement has $\dd v\ge0$ and $-f\,\dd v^2+2\,\dd v\,\dd r\le0$, i.e.\ $\dd v\,(|f|\dd v+2\,\dd r)\le0$, which is \eqref{blueshiftbound}; integrating over the annulus, or from $r=h-\varepsilon$ to $r=0$ using that $|f|\ge\Delta_\varepsilon$ on all of $(0,h-\varepsilon]$, gives the bounds on the elapsed advanced time. On the static background ($\mu=0$) $r$ is affine along outgoing radial null geodesics and $k^v=2k^r/f_0$, whence the flux bound. (b) Equation \eqref{mu_wave} is derived in Appendix~C from \eqref{massfluxeq}--\eqref{massfluxeq_u}, the Raychaudhuri equations and the conservation of each dust stream. With $r\ge r_1$ and part (a), the coefficient is at most $c_\varepsilon$, so $\partial_u(\ee^{-c_\varepsilon\mu}\partial_v\mu)=\ee^{-c_\varepsilon\mu}\,[2/(r|f|)-c_\varepsilon]\,\partial_u\mu\,\partial_v\mu\le0$. Hence $\partial_v\ee^{-c_\varepsilon\mu(u,v)}\ge\partial_v\ee^{-c_\varepsilon\mu(u_0,v)}$, and integrating in $v$ gives \eqref{apriori}, which is equivalent to \eqref{muboundB} when $B>0$; for $\mu(u_0,v_0)=0$ and boundary masses below $(\ln2)/c_\varepsilon$ each exponential exceeds $1/2$, so $B>0$. (c) The first statement is \eqref{DTRmass}, derived in Appendix~D, with $|f_S|\ge\Delta_\varepsilon$. Equation \eqref{Gamma_def} is the continuum limit of \eqref{DTRmass} for infinitesimal $\din=\Lin\dd v$ (also in Appendix~D); \eqref{deltabound} follows by integration along the portion of the trajectory in the annulus, using $\Lin\le L_{\max}$ and the bound $2(h-\varepsilon-r_1)/\Delta_\varepsilon\le2(h-\varepsilon)/\Delta_\varepsilon$ on the advanced time spent there. $\hfill\square$

Theorem~2 makes the role of the gap explicit: by \eqref{mu_wave} and \eqref{DTRmass} the coupling between the two fluxes is \emph{exactly} $2/(r|f|)$, so on the annulus $r_1\le r\le h-\varepsilon$, where $r$ is bounded above and below, it is uniformly bounded if and only if $\inf|f|>0$ there. The gap is therefore not an auxiliary hypothesis but precisely the condition that controls the in--out coupling. By contrast, the ingoing equation alone carries no such information: in the Eddington gauge $-\partial_ur/\Omega^2=1/2$ and \eqref{massfluxeq} gives $\partial_v\mu=\Lin(v)$, so a bounded influx changes the mass at most linearly in $v$ with or without an inner horizon. What distinguishes the two cases is the range of $v$ [part (a)] and the coupling to $T_{uu}$ [parts (b) and (c)], and both are controlled by $|f|$.

\begin{figure}[t]
\centering
\includegraphics[width=0.95\linewidth]{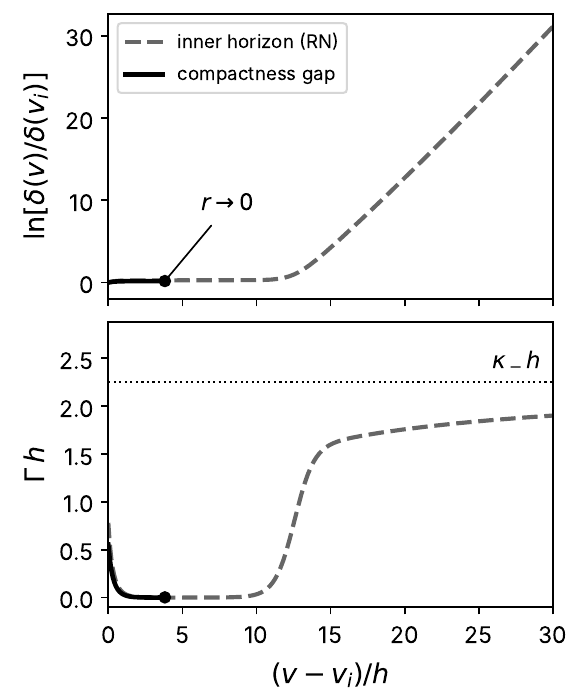}
\caption{Ori model \eqref{Gamma_def} integrated numerically for the same Price-law influx profile $\Lin\propto v^{-p}$, with the same normalization in units of $h$, on a Reissner--Nordstr\"om interior and on the exponentially screened integrable interior \eqref{q_exp} (parameters in Appendix~D; $h$ denotes the event-horizon radius of each geometry). Top: logarithmic amplification of the outgoing-shell mass. Bottom: instantaneous e-folding rate $\Gamma$. With an inner horizon (dashed curves) $\Gamma\to\kappa_-$ (dotted line; the approach is $\kappa_--p/v$) and the growth is exponential for all $v$. With a compactness gap (solid curves) $\Gamma$ decays with the influx, the amplification accumulated in the annulus $r_1\le R\le h-\varepsilon$ obeys \eqref{deltabound}, and the integration continued to $r=0$ (dot) shows that the total amplification remains of order unity after a finite advanced time.}
\label{fig:linear}
\end{figure}

The comparison with an inner horizon is instructive, and is shown in Fig.~\ref{fig:linear} for the same influx on both geometries. At an inner horizon $R\to r_-$, and the lapse on the shell, $g(v)\equiv f(v,R(v))$, obeys $\dot g=-\kappa_-g-2\Lin/r_-$ near $r_-$, with attractor $g\to-2\Lin/(\kappa_-r_-)$ (Appendix~D). Inserting this in \eqref{Gamma_def} gives $\Gamma\to\kappa_-$ \emph{independently of the amplitude of the influx}: the rate is set by the geometry, the range of $v$ is infinite, and $\delta\sim v^{-p}\ee^{\kappa_-v}$ \cite{Ori1991}. With a compactness gap there is no such attractor: $\Gamma$ is proportional to $\Lin$, vanishes with it, and acts for a finite advanced time.

\textbf{Corollary 1.} \textit{Let $\rho(r)$ satisfy (i)--(iii), and let the corresponding static interior be irradiated, from $v=v_i$ on, by ingoing null dust of luminosity $0\le\Lin(v)\le L_{\max}$. Then, for all $v\ge v_i$: (1) the exact solution is \eqref{genVaidya} with $\mu(v)=\int_{v_i}^v\Lin$, and $f(v,r)\le f_0(r)\le-\Delta_\varepsilon$ on $(0,h-\varepsilon]$, with the same $\Delta_\varepsilon$ as in \eqref{gap_def}; $f(v,\cdot)$ has no zero in $(0,h)$ and the apparent horizon lies at $r\ge h$. (2) Hypotheses (a) and (b) of Theorem~1 hold. (3) Consequently parts (a) and (c) of Theorem~2 apply with the static gap $\Delta_\varepsilon$: the advanced time spent in the gapped region is bounded by \eqref{blueshiftbound}, and an outgoing thin shell crossing the influx is amplified at most according to \eqref{Gamma_def}--\eqref{deltabound} while it lies in $r_1\le r\le h-\varepsilon$, so that no exponential mass inflation occurs there. When a continuous outgoing null flux is present as well, the decomposition \eqref{mu_def} and the gap inequality of Theorem~2(a) continue to hold, and the radiative mass is controlled by Theorem~2(b).}

\textit{Proof.} (1) For $T^a{}_b=-\rho\,\delta^a{}_b$ plus ingoing null dust the field equations reduce to $\partial_rm=4\pi r^2\rho(r)$ and $\partial_vm=\Lin(v)$ \cite{WangWu1999}, whose solution with $\mu(v_i)=0$ is \eqref{genVaidya}. Since $\Lin\ge0$, $\mu(v)\ge0$ and $f(v,r)=f_0(r)-2\mu(v)/r\le f_0(r)$, which is $\le-\Delta_\varepsilon$ on $(0,h-\varepsilon]$ by Proposition~1 and $<0$ on $(0,h)$; hence the outermost zero of $f(v,\cdot)$ is at $r\ge h$. (2) $T_{vv}\propto\Lin\ge0$ and $T_{uu}=0$; null dust does not contribute to $T_{uv}$, so the density is still the static profile, $\rho(r)\le\bar\rho_0(r)/3\le\bar\rho/3$, the last step because $m\ge m_0$. (3) The hypotheses of Theorem~2(a),(c) are (1) and (2); with an additional outflux $\Lout\ge0$, Theorem~2(a),(b) apply directly, since their proof uses only \eqref{mu_def} and the NEC. In particular the growth exponent in \eqref{deltabound} is finite and proportional to the accreted mass, whereas exponential inflation requires $\ln\delta$ to grow linearly in $v$ over an unbounded range of $v$, which \eqref{blueshiftbound} excludes. $\hfill\square$

The static observation that no second zero of $f$ exists is thus only the starting point. Theorem~1 shows that, in contrast with the slowly-evolving-horizon scenarios of \cite{CarballoRubio2022,CarballoRubio2024PRL}, the lapse cannot approach zero dynamically as long as the NEC and the screening condition hold, and Theorem~2 converts the surviving gap into explicit bounds on the quantities that grow without bound at a Cauchy horizon.

\section{Discussion}

The result of this {paper} is summarized by the chain \eqref{causal_chain}. A screened marginal profile has a compactness gap (Proposition~1); the gap is propagated by the NEC and the screening condition (Theorem~1); and the gap is the quantity that bounds, on the one hand, the blueshift factor $\dd v/\dd r=2/f$ and the advanced time available to the interior, and on the other hand the coupling $2/(r|f|)$ between ingoing and outgoing fluxes through which the \MS mass inflates (Theorem~2). An inner horizon is the limiting case $\Delta_\varepsilon\to0$, in which these bounds are lost; there the growth rate is instead fixed by the surface gravity $\kappa_-$, as derived in Appendix~D.

\section{Scope and outlook}

The analysis is formulated in spherical symmetry and for matter obeying the NEC. Theorem~1 is nonlinear and requires only its two hypotheses on the evolved domain, which hold for the fluid-plus-null-dust model; the explicit mass bounds of Theorem~2 are obtained for counter-streaming null dust and thin shells, the models in which mass inflation was established, with \eqref{muboundB} explicit for boundary data below the threshold $(\ln2)/c_\varepsilon$. The uniform estimates are formulated on the annulus $r\ge r_1$, where the static gap alone controls the coefficient $2/(r|f|)$, since $r|f_0|=2m_0-r$ vanishes at the center. This is a property of the uniform estimate rather than of the dynamics: for $\mu>0$ one has $r|f|=2m_0(r)+2\mu-r\to2\mu$ as $r\to0$, so the coupling remains finite at the center, where the accreted mass produces a Schwarzschild-like singularity \cite{ArrecheaLiberatiNeshatVellucci2025}; no infinite range of $v$ is involved. The NEC plays an essential role: for $\mu<0$, as in an evaporating interior, $f\simeq1-8\pi A+2|\mu|/r$ changes sign near the center at $r_-\simeq2|\mu|/(8\pi A-1)$, with surface gravity $\kappa_-\simeq(8\pi A-1)^2/(4|\mu|)$, so that the semiclassical evolution of these interiors is a natural next problem. In this sense the present analysis complements the rigorous nonlinear analyses of \cite{Dafermos2003,LukOh2019,DafermosLuk2025,Costa2015}: it isolates the specific geometric condition -- a lapse-degeneracy locus inside the trapped region -- that the Poisson--Israel--Ori mechanism requires, and shows that a compactness gap removes exactly that condition.

An arguably more ambitious goal would be to identify \emph{non-pathological} classes of inner horizons rather than excluding them outright, and it is worth being precise about why this is harder than the present construction. Our mechanism sidesteps the instability by removing its geometric prerequisite -- a second zero of $f$ -- entirely; a non-pathological-horizon program would instead have to allow $f(r_-)=0$ while controlling the DTR blueshift itself, for instance by forcing the inner-horizon surface gravity $\kappa_-\to0$ (an extremal-type degeneration, which trades exponential growth for a slower, possibly polynomial one) or by finding matter content for which the flux term in Eq.~\eqref{mu_wave} is dynamically suppressed near $r_-$ rather than merely bounded on a region away from it. Both routes require dynamical, not just static, control at the horizon itself -- precisely the point our compactness-gap argument avoids by construction, since $\Delta_\varepsilon$ is only claimed on $(0,h-\varepsilon]$. To our knowledge no explicit example of this second, more demanding, type is currently established, and we regard it as a natural but substantially harder open problem rather than one this {paper} resolves. Extending the present gap condition beyond spherical symmetry, to rotating interiors, and connecting it to fully nonlinear Einstein-matter evolutions, remain open directions as well. In this sense, marginally screened densities $\rho\sim A/r^2$ furnish a concrete, if narrow, class of interiors for which the standard mass-inflation channel is structurally, not just instantaneously, inoperative -- a statement compatible with, but weaker than, strong cosmic censorship itself; indeed, the regularity class in which such a statement is made matters, since the weaker $C^0$ formulation of strong cosmic censorship has recently been shown to fail for vacuum perturbations of Kerr, which admit a continuous extension across the Cauchy horizon \cite{DafermosLuk2025} -- our obstruction, like the mass-inflation mechanism itself, concerns the stronger, lapse-regularity sense, not extendibility in the weakest, merely continuous, one.

\section*{Data availability}

{There are no publicly available research data or software supporting this manuscript. Requests for further information or data should be sent to the authors.}

\appendix

\section{Appendix A: Uniform convergence of the marginal profile}
\label{app:convergence}

Here we make precise the sense in which $\rho(r)=A/r^2+o(r^{-2})$ guarantees the integrability condition \eqref{integrable_def}, addressing the convergence concern directly. Write
\begin{equation}
r^2\rho(r) = A + \eta(r),\qquad \eta(r)\to 0 \ \text{as } r\to0^+,
\label{app_eta_def}
\end{equation}
with $\eta$ continuous on {$(0,r_c]$ for some $r_c>0$}. On this interval, by \eqref{R_rho},
\begin{equation}
r^2R(r) = 8\pi\left[4r^2\rho(r)+r^3\rho'(r)\right],
\end{equation}
where, from \eqref{app_eta_def}, $r^3\rho'(r) = -2A - 2\eta(r) + r\,\eta'(r)$, so that $r^2R(r)=8\pi\,[2A+2\eta(r)+r\eta'(r)]$. Assuming, as is standard for a marginal profile, that $\eta$ is $C^1$ with $r\eta'(r)\to0$ as $r\to0^+$ (i.e. the correction is differentiably subleading, not merely bounded), $r^2R(r)\to16\pi A$ and there exists $K>0$, independent of $r$, such that
\begin{equation}
r^2|R(r)| \le 8\pi K
\qquad (0<r\le {r_c}).
\label{app_uniform_bound}
\end{equation}
Since $r^2|R(r)|$ is continuous and bounded on $(0,{r_c}]$, it is integrable there, and Eq.~\eqref{app_uniform_bound} gives
\begin{equation}
\int_0^{{r_c}} r^2|R(r)|\,\dd r \le 8\pi K\,{r_c} <\infty,
\end{equation}
which is \eqref{integrable_def}. This is the precise sense in which the marginal profile \eqref{marginal_profile} produces an integrable singularity: it is not merely that $r^2R(r)$ has a limit, but that this limit is approached uniformly, which is what allows the bound $K$ in \eqref{app_uniform_bound} to be taken independent of $r$ and hence what justifies passing the limit through the integral in Eq.~\eqref{ricci_conv_integral} of the main text.

\section{Appendix B: Rational screener}
\label{app:rational}

As a second, algebraically independent example of a marginally screened profile, consider
\begin{equation}
\rho_{\mathrm{rat}}(r) = \frac{A}{r^2\,(1+r/b)^{2}},\qquad A,b>0.
\label{app_rho_rat}
\end{equation}
Near the origin $\rho_{\mathrm{rat}}(r)=A/r^2+o(r^{-2})$, so \eqref{app_rho_rat} is marginal in the same sense as \eqref{marginal_profile}. Using Eq.~\eqref{mass_eq},
\begin{equation}
m_{\mathrm{rat}}(r) = 4\pi A\,\frac{r}{1+r/b} = \frac{4\pi A b\,r}{b+r},
\label{app_m_rat}
\end{equation}
so that, with $x\equiv r/b$,
\begin{equation}
q_{\mathrm{rat}}(r) = \frac{m_{\mathrm{rat}}(r)}{r} = \frac{4\pi A}{1+x}.
\label{app_q_rat}
\end{equation}
This is manifestly monotonically decreasing in $x$ for all $x>0$, with $q_{\mathrm{rat}}(0^+)=4\pi A$ and $q_{\mathrm{rat}}(r)\to0$ as $r\to\infty$; in fact $q_{\mathrm{rat}}'(r)<0$ holds identically, without need of an inequality analogous to \eqref{qprime_exp_negative}. Choosing $A>1/(8\pi)$ places the center in the trapped region as before, $q_{\mathrm{rat}}(h)=1/2$ defines a unique event horizon at $h=b(8\pi A-1)$, non-degenerate with $\kappa_+=1/(16\pi A\,b)$, and
\begin{equation}
\Delta_\varepsilon^{(\mathrm{rat})} = \frac{8\pi A}{1+(h-\varepsilon)/b}-1 >0
\end{equation}
for every $\varepsilon\in(0,h)$, exactly as in the exponential case. The Ricci scalar again behaves as $R(r)\sim 1/r^2$ near the origin by the same argument as in Appendix~A, so the singularity is integrable. This second, algebraically simpler example confirms that the existence of a compactness gap is not a special feature of the exponential screening \eqref{q_exp}, but persists across at least two structurally different marginal profiles.

\section{Appendix C: Double-null identities and the equation for the radiative mass}
\label{app:doublenull}

For the metric \eqref{doublenullmetric} with $G_{\mu\nu}=8\pi T_{\mu\nu}$, the $uu$ and $vv$ equations are the Raychaudhuri equations
\begin{equation}
\partial_u\!\left(\frac{\partial_ur}{\Omega^2}\right)=-\frac{4\pi r}{\Omega^2}T_{uu},\qquad
\partial_v\!\left(\frac{\partial_vr}{\Omega^2}\right)=-\frac{4\pi r}{\Omega^2}T_{vv},
\label{app_raych}
\end{equation}
and the $uv$ equation is $r\,\partial_u\partial_vr+\partial_ur\,\partial_vr+\Omega^2/4=4\pi r^2T_{uv}$. Differentiating $m=(r/2)(1+4\,\partial_ur\,\partial_vr/\Omega^2)$ and using these three equations gives \eqref{massfluxeq}--\eqref{massfluxeq_u}. The identity \eqref{mono_v} follows from $\partial_v(2m/r)=(2/r)\,\partial_vm-(2m/r^2)\,\partial_vr$ upon inserting \eqref{massfluxeq} and $m/r-4\pi r^2\varrhoeff=4\pi r^2(\bar\rho/3-\varrhoeff)$; \eqref{mono_u} is obtained by exchanging $u$ and $v$. Equations \eqref{massfluxeq}--\eqref{mono_u} are covariant under $u\to U(u)$, $v\to V(v)$.

\emph{Static solutions.} For $f<0$ let $\dd r_*=\dd r/f$, $u=r_*-t$, $v=r_*+t$. Then \eqref{metric_static} takes the form \eqref{doublenullmetric} with $\Omega^2=|f|$ and $\partial_ur=\partial_vr=f/2<0$; $T_{ab}=-\rho\,g_{ab}$ in the $(u,v)$ block gives $T_{uu}=T_{vv}=0$ and $T_{uv}=\Omega^2\rho/2$, so \eqref{mono_v} reduces to $f\,q'=-4\pi r f(\bar\rho/3-\rho)$, i.e.\ to \eqref{qprime2}. In this (Eddington) gauge $\Omega^2=|f|\to0$ at a zero of $f$: this is the lapse degeneracy on which the standard mechanism relies.

\emph{Conservation.} For $T^{\mu\nu}=-\rho\,h^{\mu\nu}+p_t\,s^{\mu\nu}$, with $h$ and $s$ the projections on the $(u,v)$ quotient and on the spheres, $\nabla_\mu T^{\mu}{}_{a}=0$ gives $\partial_a\rho+(2/r)(\rho+p_t)\,\partial_ar=0$, hence $\dd\rho\wedge\dd r=0$ and $\rho=\rho(r)$ wherever $\dd r\neq0$. For non-interacting null dust, $\partial_u(r^2T_{vv})=0$ and $\partial_v(r^2T_{uu})=0$, so $4\pi r^2T_{vv}=\Lin(v)$ and $4\pi r^2T_{uu}=\Lout(u)$.

\emph{Equation \eqref{mu_wave}.} With $T_{uv}=\Omega^2\rho(r)/2$, Eq.~\eqref{massfluxeq} reads $\partial_vm=m_0'(r)\,\partial_vr+\partial_v\mu$, with
\begin{equation}
\partial_v\mu=\Lin(v)\,{\chi_u},\qquad \partial_u\mu=\Lout(u)\,{\chi_v},
\label{app_muv}
\end{equation}
where ${\chi_u}\equiv-2\,\partial_ur/\Omega^2>0$ and ${\chi_v}\equiv-2\,\partial_vr/\Omega^2>0$ in a trapped region. By \eqref{app_raych}, $\partial_u{\chi_u}=2\Lout/(r\Omega^2)$, and by \eqref{MSdouble}, ${\chi_u\chi_v}=-f/\Omega^2=|f|/\Omega^2$. Therefore
\begin{equation}
\partial_u\partial_v\mu=\frac{2\Lin\Lout}{r\,\Omega^2}=\frac{2\,\Lin {\chi_u}\,\Lout {\chi_v}}{r\,|f|}=\frac{2\,\partial_u\mu\,\partial_v\mu}{r\,|f|},
\end{equation}
which is \eqref{mu_wave}. Reissner--Nordstr\"om belongs to the same class of sources, with $\rho=Q^2/(8\pi r^4)$ and $m_0=M-Q^2/(2r)$, and for it \eqref{mu_wave} is the wave equation for the mass function, with a source quadratic in the fluxes, that underlies the analysis of \cite{PoissonIsrael1990}. We have verified \eqref{massfluxeq}--\eqref{mono_u}, \eqref{app_raych} and \eqref{mu_wave} by direct symbolic computation of the Einstein tensor. Note that in the Eddington gauge of the static background ${\chi_u}=1$, so $\partial_v\mu=\Lin(v)$: this is the statement that a bounded influx changes the mass linearly in $v$, used in the main text.

\section{Appendix D: DTR relation, Ori limit and numerical illustration}
\label{app:DTR}

\emph{DTR relation.} Let an ingoing shell at $v=v_0$ and an outgoing shell at $u=u_0$ cross at a sphere of radius $r_0$, in continuous double-null coordinates. The derivative $\partial_vr$ is tangential to the outgoing shell and is therefore continuous across $u=u_0$, while $\partial_ur$ is continuous across $v=v_0$. Denoting by $(\alpha,\alpha')$ the values of $\partial_vr$ for $v<v_0$ and $v>v_0$, and by $(\beta,\beta')$ those of $\partial_ur$ for $u<u_0$ and $u>u_0$, Eq.~\eqref{MSdouble} gives at the crossing $f_S\propto\alpha\beta$, $f_E\propto\alpha'\beta$, $f_W\propto\alpha\beta'$, $f_N\propto\alpha'\beta'$ with the same factor $-4/\Omega^2$, hence $f_Nf_S=f_Ef_W$ \cite{DrayTHooft1985,Redmount1985,BarrabesIsrael1991}. The relation involves only the \MS masses at $r_0$ and holds for any spherically symmetric sectors. Writing $f_i=1-2m_i/r_0$, $m_E=m_S+\din$ and $m_W=m_S+\dout$,
\begin{align}
f_N&=\frac{(f_S-2\din/r_0)(f_S-2\dout/r_0)}{f_S}\nonumber\\
&=f_S-\frac{2(\din+\dout)}{r_0}+\frac{4\din\dout}{r_0^2f_S},
\end{align}
which is \eqref{DTRmass}. In a trapped region $\din,\dout\ge0$ under the NEC by \eqref{massfluxeq}--\eqref{massfluxeq_u}.

\emph{Ori limit.} Let the outgoing shell, of mass $\delta\equiv\dout$, sit at $r=R(v)$ in the geometry \eqref{genVaidya}, so that $\dd R/\dd v=f(v,R)/2$, and let the influx between $v$ and $v+\dd v$ act as an ingoing shell with $\din=\Lin\dd v$. The mass of the outgoing shell after the crossing is $m_N-m_E=\delta\,[1+2\din/(R|f|)]$ by \eqref{DTRmass}, i.e.
\begin{equation}
\frac{\dd\delta}{\dd v}=\frac{2\,\Lin(v)}{R\,|f(v,R)|}\,\delta ,
\end{equation}
which is \eqref{Gamma_def}; applied to a Reissner--Nordstr\"om background it reproduces the growth law of \cite{Ori1991} (see below). The static profile $m_0(r)$ is the same on both sides of the shell and drops out of $\delta$.

\emph{Inner-horizon attractor.} Along the shell, $g(v)\equiv f(v,R(v))$ obeys $\dot g=\tfrac12\,\partial_rf\,g-2\Lin/R$. If $f(v,\cdot)$ has an inner horizon, $\partial_rf\to-2\kappa_-$ and $R\to r_-$, so $\dot g=-\kappa_-g-2\Lin/r_-$, whose solutions relax at the rate $\kappa_-$ to $g\simeq-2\Lin/(\kappa_-r_-)$ for slowly varying $\Lin$. Then $\Gamma=2\Lin/(R|g|)\to\kappa_-$, and including the first correction for $\Lin\propto v^{-p}$, $\Gamma\simeq\kappa_--p/v$, i.e.\ $\delta\propto v^{-p}\ee^{\kappa_-v}$. With a compactness gap $|g|\ge\Delta_\varepsilon$ and no such attractor exists.

\emph{Numerical illustration.} {Figure~\ref{fig:linear} is obtained by integrating $\dot R=g/2$, the equation for $g$ above, and Eq.~\eqref{Gamma_def}, in units of the event-horizon radius $h$ of each background, for $\Lin=\mu_\infty(p-1)v_i^{\,p-1}v^{-p}$ with $p=12$, $v_i=5h$, and total accreted mass $\mu_\infty=10^{-2}h$, the shell being launched at $R(v_i)=0.9\,h$. The backgrounds are Reissner--Nordstr\"om with $Q/M=0.9$ and the exponentially screened profile \eqref{q_exp} with $4\pi A=3/2$. The integration was performed with a standard adaptive solver for stiff ordinary differential equations, and the results are insensitive to the numerical tolerances. In the first case the rate approaches the surface gravity of the final inner horizon, $\kappa_-h\simeq2.25$, as $\kappa_--p/v$. In the second case, with $\varepsilon=0.1\,h$ and $r_1=0.05\,h$ (so that $\Delta_\varepsilon\simeq0.09$), the lapse on the shell remains above $\Delta_\varepsilon$ throughout the annulus, and the shell reaches $r=0$ after $\Delta v\simeq3.8\,h$ with $\ln[\delta/\delta(v_i)]\simeq0.19$, accumulated almost entirely inside the annulus, in agreement with Eqs.~\eqref{blueshiftbound} and \eqref{deltabound}.}

\bibliographystyle{apsrev4-1}
\bibliography{revised1}

\end{document}